\documentclass[epj]{svjour}
\usepackage[pdftex]{graphicx}
\usepackage[latin1]{inputenc}
\usepackage{here}
\usepackage{url}
\begin{document}
\title{A Hot Cavity Laser Ion Source at IGISOL}
\author{M. Reponen \inst{1} \and T. Kessler\inst{1} \and I.D. Moore\inst{1} \and S. Rothe\inst{2} \and J. Äystö\inst{1}
}                     
\offprints{M. Reponen}          
\institute{Department of Physics, University of Jyväskylä, PO Box 35 (YFL), FI-40014, Finland \and AG Larissa/Quantum, Institut für Physik, Johannes Gutenberg Universität, Staudinger Weg 7, D-55128 Mainz, Germany}
\date{Received: date / Revised version: date}
%
\abstract{A development program is underway at the IGISOL (Ion Guide Isotope Separator On-Line)
facility, University of Jyväskylä, to efficiently and selectively produce low-energy radioactive ion beams of silver isotopes and isomers, with a particular interest in \textit{N=Z} $^{94}$Ag. A hot cavity ion source has been installed, based on the FEBIAD (Forced Electron Beam Induced Arc Discharge) technique, combined with a titanium:sapphire laser system for selective laser ionization. The silver recoils produced via the heavy-ion fusion-evaporation reaction, $^{40}\textrm{Ca}\left(^{58}\textrm{Ni},\textrm{p3n}\right)$$^{94}\textrm{Ag}$, are stopped in a graphite catcher, diffused, extracted and subsequently ionized using a three-step laser ionization scheme. The performance of the different components of the hot cavity laser ion source is discussed and initial results using stable $^{107,109}$Ag are presented.
}

%
\maketitle
\section{Introduction}
\label{intro}
The nuclear structure study of radioactive neutron-deficie\-nt silver isotopes in the region of the \textit{N=Z} line has been of considerable interest for several years. The (21$^+$) isomeric state of the lightest known isotope of silver, \textit{N=Z} $^{94}$Ag, has been identified as a spin trap having the highest spin ever observed for $\beta$-decaying nuclei. The isomer is characterised by a half-life of 0.39(4) s \cite{mukha04}, a high excitation energy of 6.7(5) MeV \cite{mukha06} and a high spin \cite{Plett04}. These unique properties are matched by several decay modes including $\beta$ decay, one-proton and two-proton emission which, if confirmed, would make this state unprecedented in the entire known Segré chart. Additionally, it was discussed by Mukha \textit{et al} that the unexpectedly large probability for the enhanced two-proton decay could be attributed to a strongly deformed prolate shape in $^{94}$Ag \cite{mukha06}. Large-scale shell model calculations performed by Kaneko \textit{et al} \cite{kaneko08} do not support such an observation and only a direct measurement of the nuclear quadrupole moment of the isomer will clarify this discrepancy.

Recently, the existence of the two-proton decay mode was questioned \cite{pechenaya08}. Furthermore, mass measurements on isotopes in the vicinity of $^{94}$Ag have been performed which, when combined with the original spectroscopic decay data, lead to a puzzling contradiction. The deduced mass-excess values for the isomer differ by approximately 1.4 MeV using either the one-proton or two-proton decay data \cite{anu08}. To solve this puzzle, direct mass measurements of $^{94}$Ag and $^{94}$Ag $(21^+)$ are needed which will afford a unique determination of the energy of these states. This poses a challenge for the production of such exotic species. At the IGISOL facility, University of Jyväskylä, a development program to selectively and efficiently produce a low-energy ion beam of the isotope/isomer of interest has been initiated \cite{kessler08}. In addition to performing high-precision mass measurements, we intend to perform in-source resonance ionization spectroscopy on both $^{94}$Ag and $^{96}$Ag which will provide model-independent information on the change of the mean-square charge radii and, furthermore, by measuring the hyperfine structure of the isomeric states the spectroscopic quadrupole moment and thus the shape of the isomers will be determined.

In our previous article we discussed the design of a graphite catcher device similar to that proposed in \cite{kirch81}, directly heated using a primary beam of $^{40}$Ca$^{8+}$ from the JYFL K-130 cyclotron, degraded in energy to 150 MeV and with a beam intensity of 125 pnA. A maximum heating power of $\sim$19 W was provided to the first catcher foil using this method, sufficient only to release surface ions of $^{39}$K. This power is about one order of magnitude lower compared to electron-bombardment heating by a cathode in which up to 400 W can be achieved \cite{kirch75}. The first off-line tests for the development of a laser resonant ionization scheme for Ag were performed using a combination of a frequency-doubled dye laser for the first excitation step and a frequency-doubled Ti:Sapphire laser for the second excitation step. Photons from both excitation steps were used in a final non-resonant ionization step. In the present work we present a more efficient laser setup for the ionization of Ag and systematic studies of an electron-bombardment cathode-catcher hot cavity.

\section{Experimental set-up}
\label{sec:1}
The experimental set-up consists of two independent parts, the laser system and the hot cavity source.  When coupled together to form a hot cavity laser ion source, fusion-evaporation reaction products are efficiently stopped, evaporated and subsequently ionized using resonant laser ionization. The hot cavity is a Forced Electron Beam Induced Arc Discharge (FEBIAD)-type oven which utilizes electron bombardment to reach temperatures in excess of 2400$^\circ$C \cite{kirch81}. The laser system consists of two solid state Ti:Sapphire lasers pumped by a high-repetition rate (10 kHz) Nd:YAG laser and a separate Copper Vapour laser (CVL) which together provide the ionization scheme for silver, with two resonant excitation steps and one non-resonant ionization step. More details on the laser setup can be found in Moore \textit{et al} \cite{moore05}. In the following sub-sections we concentrate on the hot cavity and the coupling of the laser beams into the IGISOL target chamber.

\subsection{FEBIAD-type hot cavity}
\label{sec:2}

\begin{figure}
	\centering
		 \includegraphics[width=8 cm]{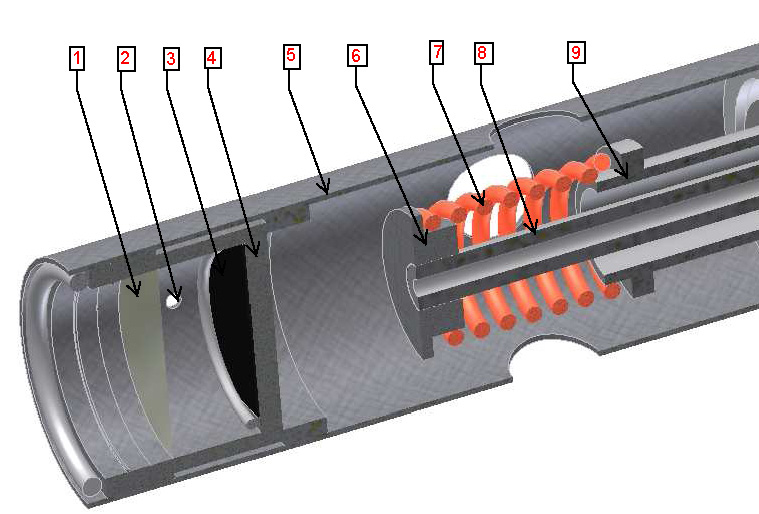}
	\caption{The FEBIAD hot cavity. 1)  2 $\mu$m tantalum foil, 2) exit nozzle,  3) graphite catcher, 4) catcher cup (anode), 5) tantalum holder, 6) tantalum cap (cathode), 7) tungsten filament, 8) inner tube, 9) outer tube.}
	\label{fig: oven}
\end{figure}

Rather than using direct ohmic heating, the FEBIAD-type hot cavity is heated by the bombardment of electrons produced from a hot filament. The specific model of the cavity used at Jyväskylä is a FEBIAD-E source \cite{kirch81}. In this version, the catcher cup (anode) and the graphite catcher are on the same electric potential unlike that of a FEBIAD-D source in which the heating of the catcher is independently biased. The principle design of the oven is presented in Fig. \ref{fig: oven}.

The oven is operated by providing a sufficiently large heating current ($\sim$25 A) to the tungsten filament from a current-stabilized power supply. This causes the tantalum cap (cathode) to emit electrons which are subsequently accelerated into the catcher cup with a bombardment voltage $V_B$. As the work function of tantalum is lower than that of tungsten, most of the electrons are emitted from the well-defined tantalum surface as opposed to the tungsten filament. The bombardment voltage is provided with an ordinary variac supply rather than a more sophisticated current-regulated supply \cite{kirch75}. 


\subsection{Hot cavity laser ion source}
\label{sec:3}

\begin{figure}
\centering
\includegraphics[width=7.5 cm]{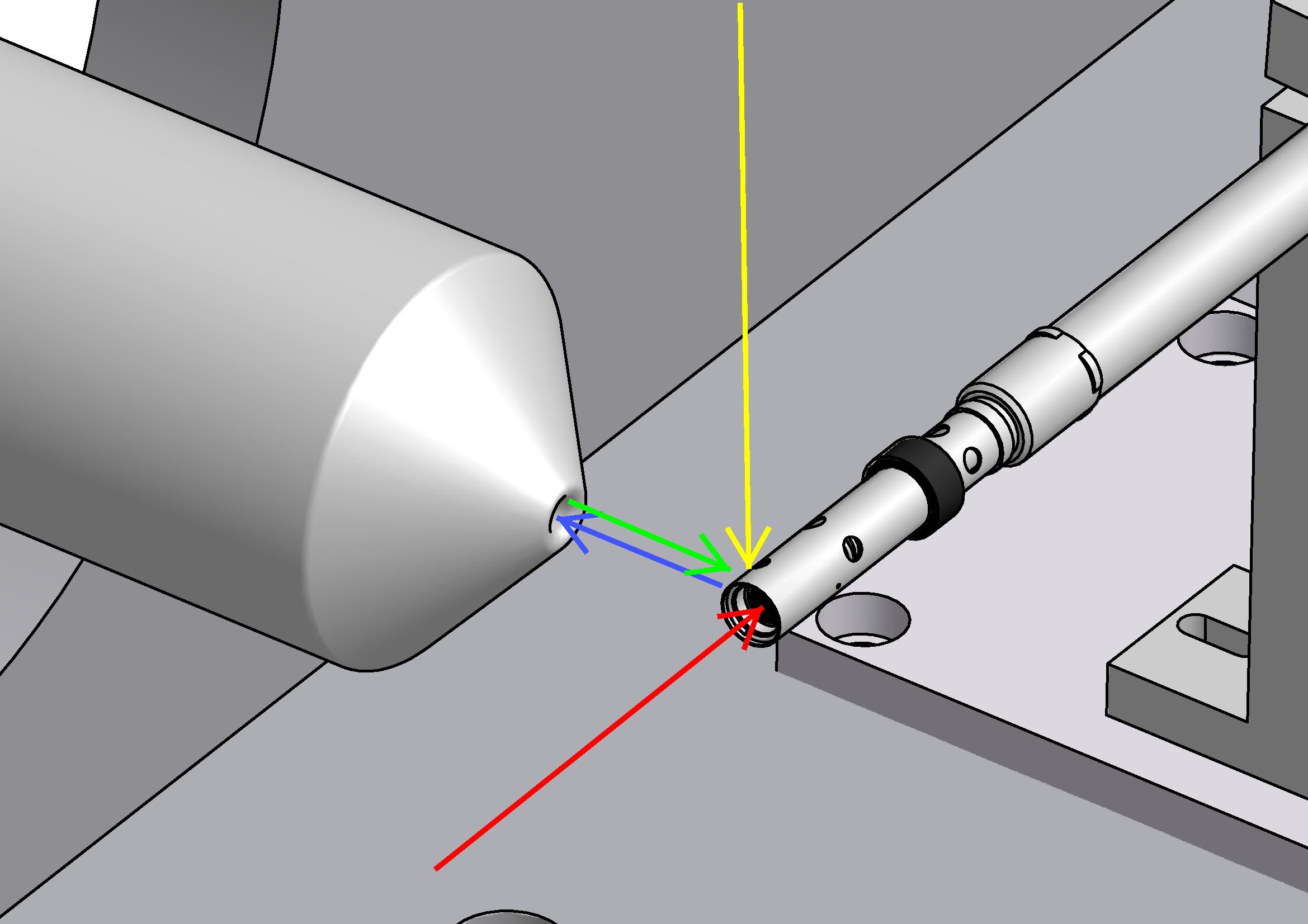}
\caption{The hot cavity laser ion source (colour on-line). The arrows are coloured as follows: red; reaction products, blue; extracted ions, green; counter-propagating laser ionization, yellow; crossed-beam laser ionization. The water cooled jacket and the target system are not displayed.}
\label{fig: hclis}
\end{figure}
A schematic illustration of the hot cavity laser ion source is shown in Fig. \ref{fig: hclis}. The hot cavity is placed inside the IGISOL target chamber which, rather than being pumped through the usual 300 mm diameter, 2.0 m long stainless steel tube to the Roots array, is pumped via the extraction chamber using a 8000 l/s diffusion pump. With this arrangement, a vacuum pressure of 10$^{-5}$ mbar can be reached inside the target chamber. The necessary current and voltage connectors for the ion source remain outside the vacuum for easy access. The set-up allows for a measurement of the electron bombardment current $I_B$ via the bombardment voltage supply and a direct measurement of the catcher plate temperature using a type C thermocouple. 


The laser beams are sent into the IGISOL target chamber in such a way that they overlap with the silver atoms effusing out of the exit hole (0.7 mm in diameter) of the hot cavity. In principle, the IGISOL mass separator geometry enables laser ionization in both a crossed and counter-propagating mode, however with a poor transmission thro\-ugh the window of the mass separator dipole magnet \newline ($\sim$10$\%$) this latter mode is presently unusable for on-line laser ionization applications. Following ionization, the ions are accelerated with a conical extraction electrode of 7 mm aperture diameter (Fig. \ref{fig: hclis}). The ion source is placed on a high voltage potential of 30 kV and the extraction electrode, at a distance of approximately 30 mm from the source, sits at a potential of $\sim$20 kV. After mass separation, the ions are detected on a set of micro-channel plates (MCP) downstream from the IGISOL focal plane.

We note that the IGISOL target chamber was not originally designed to be coupled with a high temperature ion source, and initially a number of problems occured. One minor issue was a general contamination of the target chamber from material emitted from the surface of the cavity. The primary problem however was maintaining the high voltage while operating the cavity at the highest temperatures. The extraction electrode became charged causing sparking and voltage instabilities. To address these problems a water-cooled copper jacket was constructed around the hot cavity to contain the emitted radiation, shielding both the extractor and the target chamber in general. The jacket also provides cooling for the back end of the hot cavity reducing the cooling time of the oven after use.



\section{Results}
\label{sec:4}
\subsection{Laser ionization scheme}
\label{sec:5}
\begin{figure}
	\centering
		 \includegraphics[width=7.5 cm]{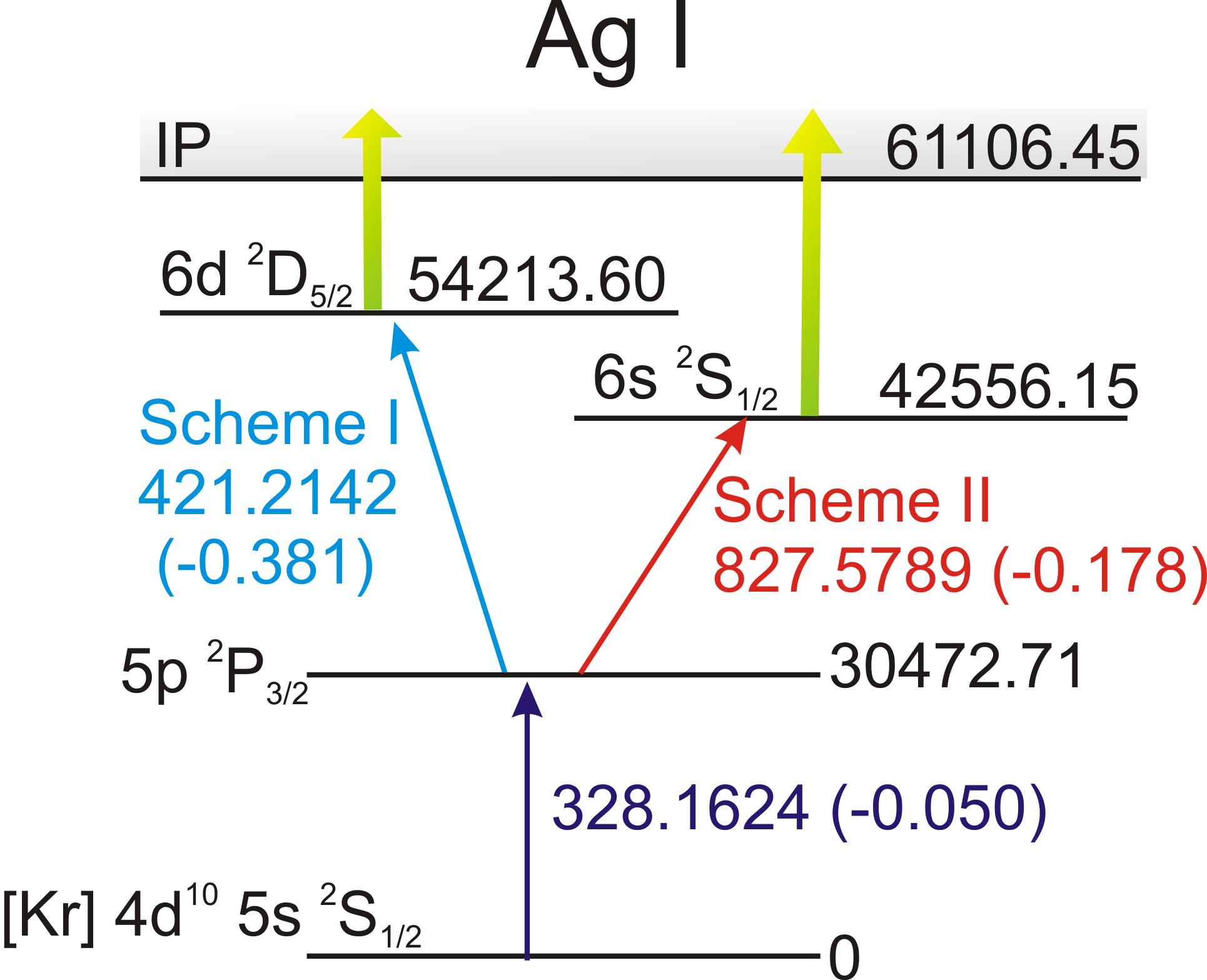}
	\caption{Investigated ionization schemes for silver. The atomic excitation energy of the levels labeled on the right is given in units of cm$^{-1}$. The wavelengths of the transitions are denoted in units of nm and are given in vacuum. Next to the wavelength, the log-gf (transition strength) is listed \cite{kuruzuc}.}
	\label{fig: scheme}
\end{figure}

An atomic beam unit, described in \cite{tordoff06}, was used to investigate the two resonant ionization schemes shown in Fig. ~\ref{fig: scheme}. Two Ti:Sapphire lasers provide the fundamental resonant wavelengths for both scheme I and scheme II. In both schemes, the first excitation step is realised using frequency-tripled laser light from a higher harmonic generation (HHG) unit. A frequency-doubled Ti:Sapphire laser provides the second resonant step in scheme I whereas a fundamental IR transition is used in scheme II. The final non-resonant ionization step in both schemes is provided by the 511 nm component of the CVL radiation. We note that scheme II has been successfully used to demonstrate not only elemental or isotopic selectivity, but also the separation of nuclear spin isomers in $^{105}$Ag at the TRIUMF Resonant Ionization Laser Ion Source (TRILIS) of the ISAC facility, TRIUMF \cite{geppert08}.

\begin{figure*}
	\centering
		 \includegraphics[width=15 cm]{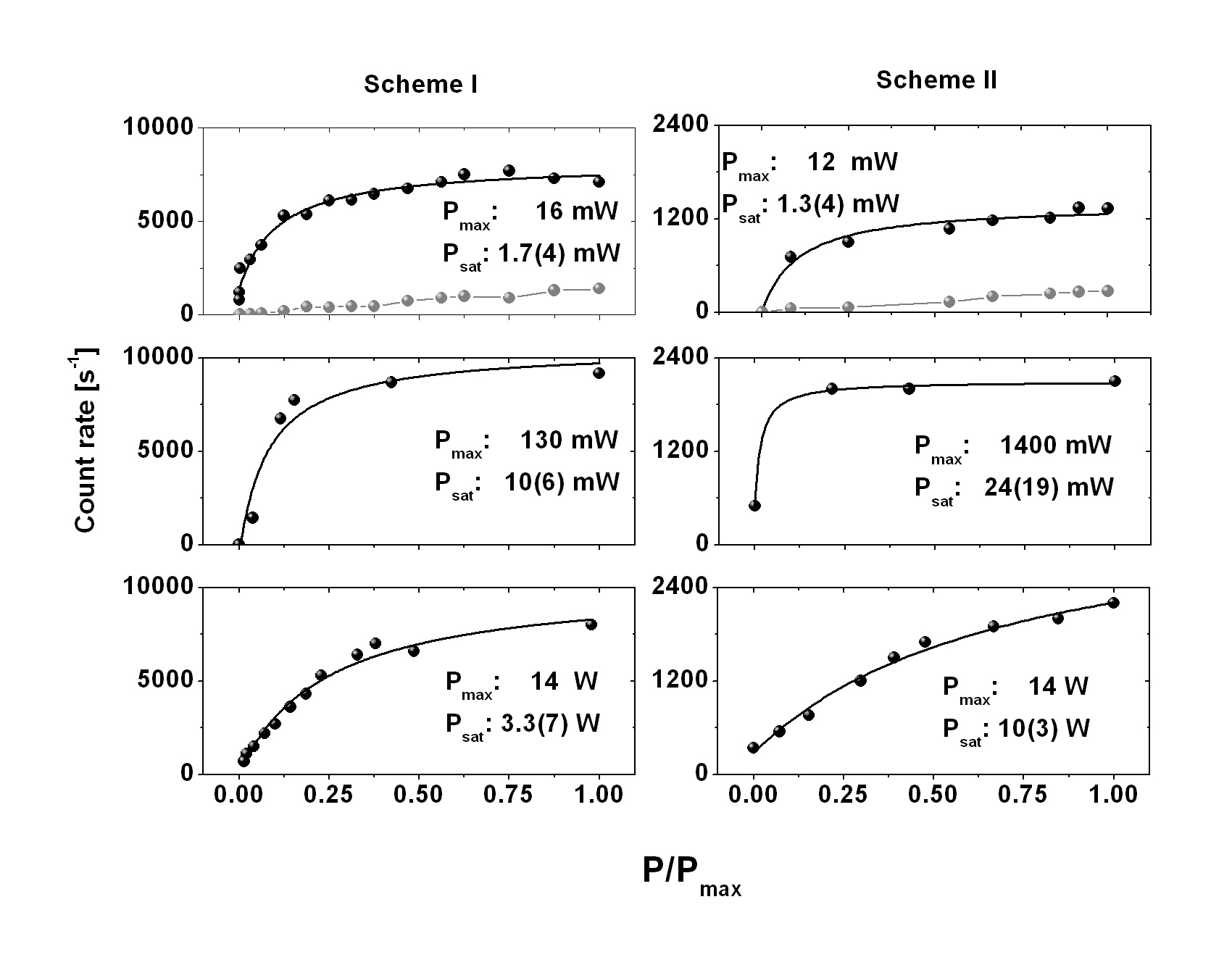}
	\caption{Comparison of the saturation of the two different laser ionization schemes (colour on-line). The top panels represent the first step, the middle panels the second step and the bottom panels the final non-resonant step.}
	\label{fig: scheme-satur}
\end{figure*}

A comparison of the saturation levels for the two schem\-es is shown in Fig. \ref{fig: scheme-satur}. The saturation data has been plotted as a function of the fraction of the maximum laser power available in each step. The first resonant excitation was corrected for the signal arising from non-resonant UV-UV ionization, indicated in the appropriate panel. It can be seen that in both ionization schemes a full saturation of the resonant steps was achieved. One can observe that even with a lower laser power and reduced oscillator strength for the second resonant step of scheme I, a considerably better saturation of the non-resonant step into the continuum was achieved compared to that involving the IR step of scheme II. Furthermore, the low saturation power required for the second step of scheme II of 24(19) mW indicates that the final non-resonant step is rather inefficient resulting in a reduced overall ionization efficiency for this scheme. Consequently, scheme I has been chosen for all future work.

\subsection{Hot cavity characteristics}
\label{sec:6}

Several systematic experiments were conducted with the hot cavity in order to determine the performance characteristics and reliability of the source. The initial tests were performed in the atomic beam unit which is rather more accessible than the IGISOL target chamber. Following this, the cavity was installed in the IGISOL chamber providing a more realistic environment.

\begin{figure}[h!]
	\centering
		 \includegraphics[width=7.5 cm]{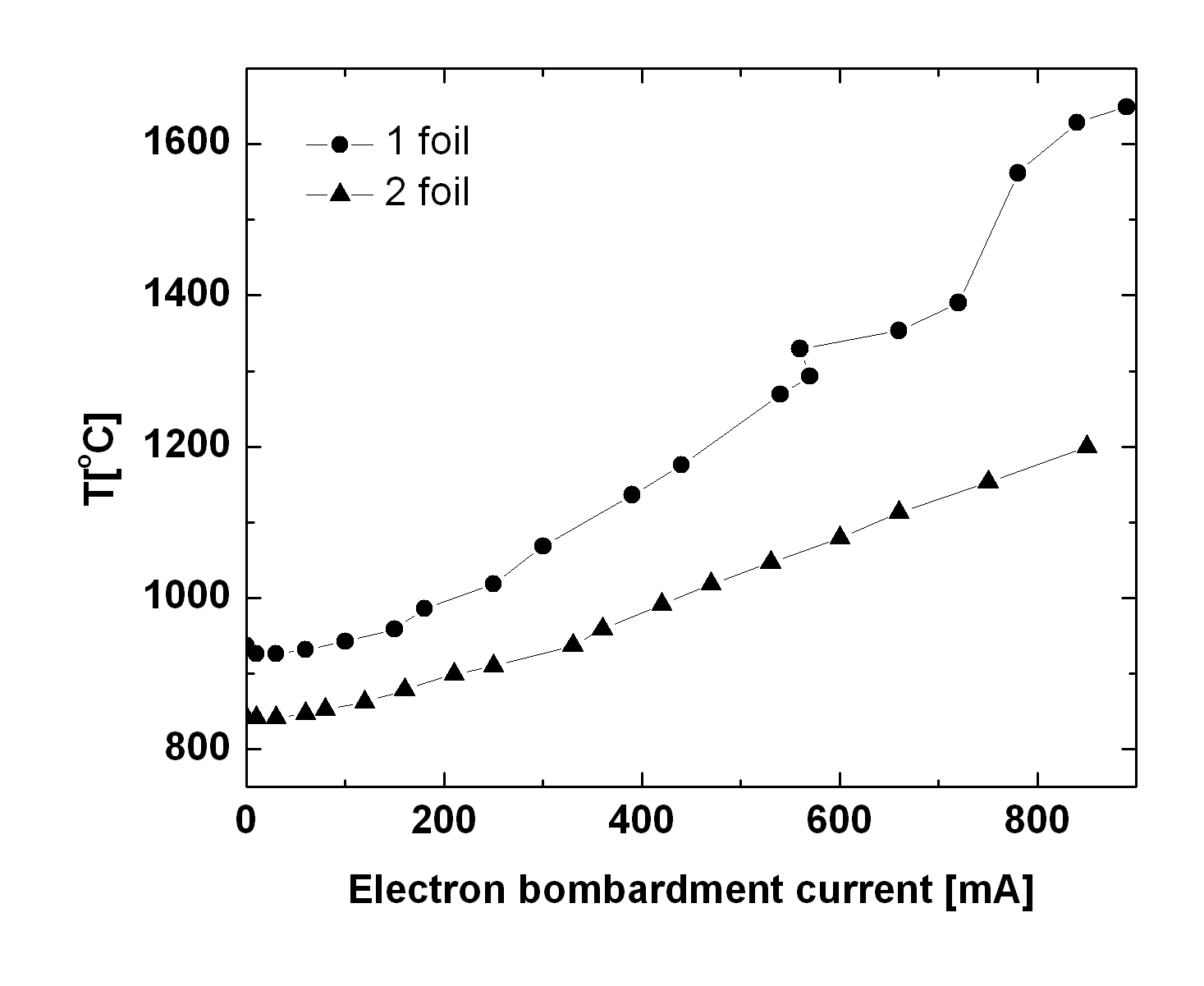}
	\caption{Effect of tantalum heat shielding on the cavity temperature. }
	\label{fig: heats}
\end{figure}

The temperature that is achievable with the hot cavity is highly dependent on the tantalum heat shielding which can be wrapped around the main body. Figure \ref{fig: heats} shows the direct effect on the temperature, measured by inserting the thermocouple into the exit hole of the cavity, when using either one or two tantalum foils of shielding. The foils are 15 $\mu$m thick and perforated to provide a small spacing between different layers. Although the temperature reaches approximately 1700$^\circ$C with only two layers of tantalum, the final set-up of the cavity features up to six layers as described in \cite{kirch81}. The cold spot of the setup remains the beam window side. To reduce heat loss, an additional three layers of graphite foil ($\sim$20 $\mu$g/cm$^{2}$) and/or a tantalum shield of 600 $\mu$g/cm$^{2}$ can be stacked in front of the catcher device \cite{kirch81}.

\begin{figure}[ht!]
	\centering
		 \includegraphics[width=7.5 cm]{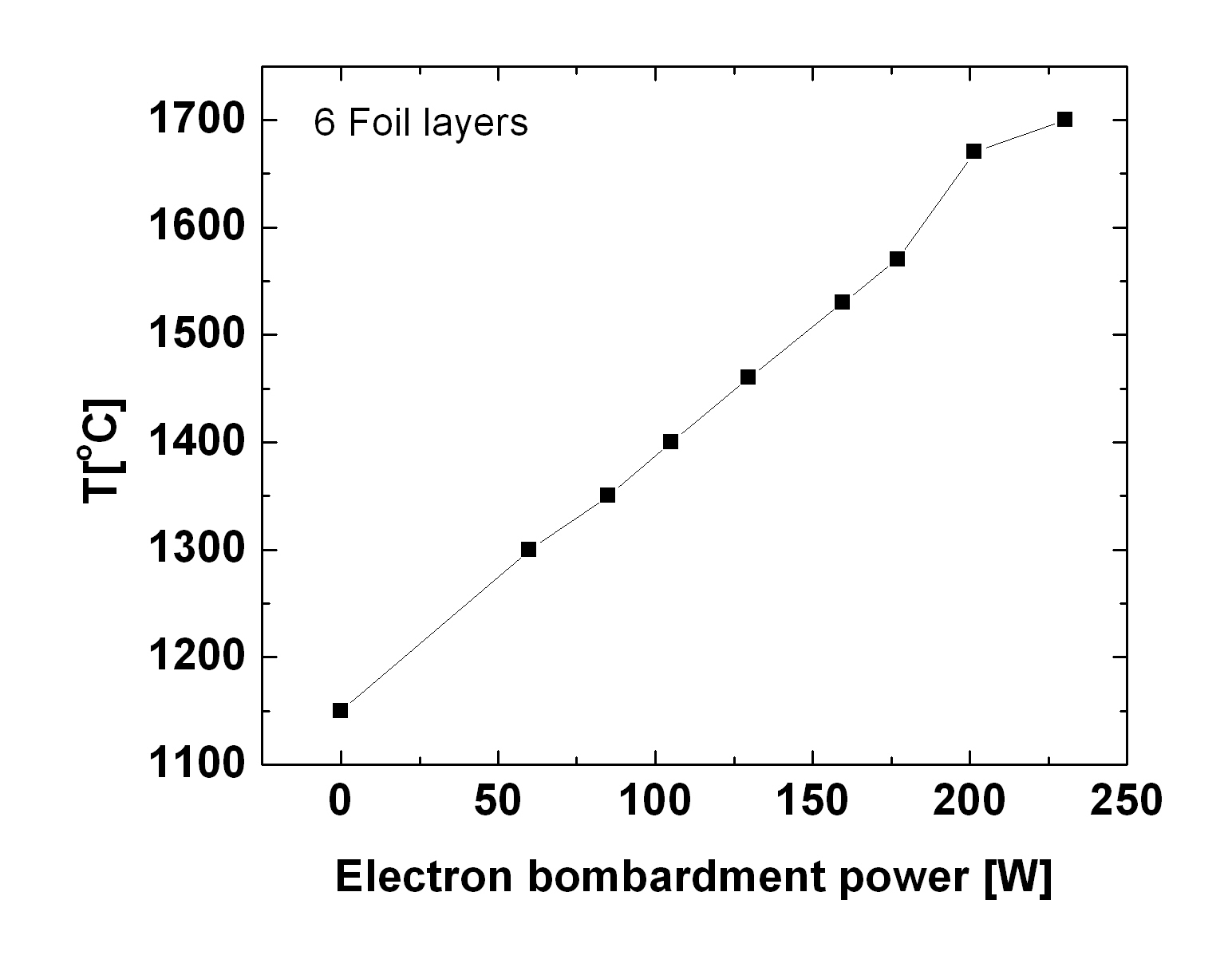}
	\caption{Catcher temperature as a function of electron bombardment power.}
	\label{fig: tp}
\end{figure}

\begin{figure}
	\centering
		 \includegraphics[width=7.5 cm]{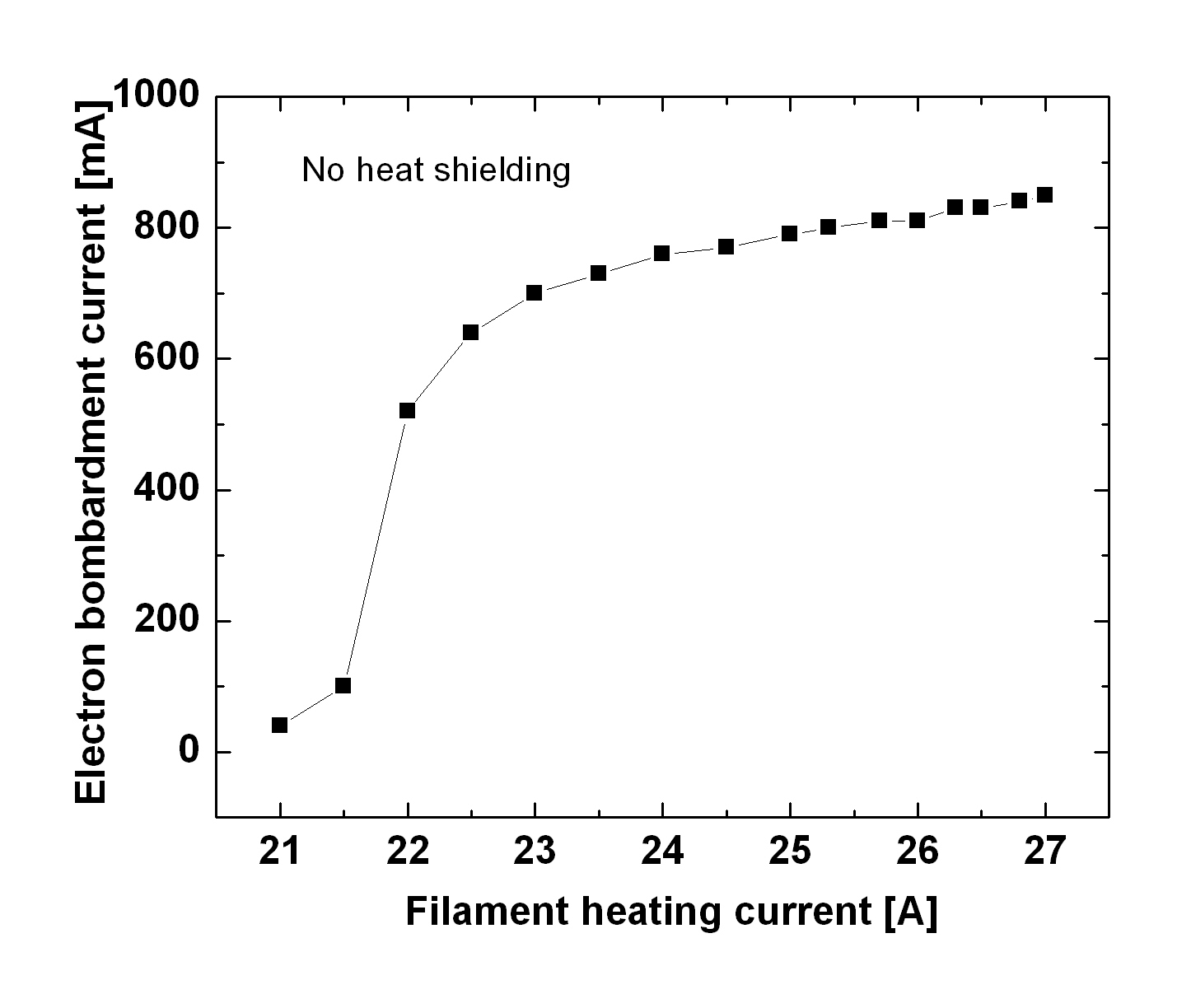}
	\caption{Electron bombardment current as a function of the filament heating current. The electron bombardment voltage was set to 340 V. The saturation effect in the bombardment current is due to  space-charge effects which occur at high current densities \cite{kirch75}.}
	\label{fig: saturationc}
\end{figure}

Using the heating of the tungsten filament alone, the catcher temperature reaches 1100$^\circ$C (with the full six layers of tantalum shielding added). This is sufficient for off-line evaporation of metallic silver placed within the cavity, however insufficient to achieve the fast diffusion times required for the extraction of short-lived isotopes/isomers of interest that are embedded within the graphite catcher. By applying a potential between the cathode and the catcher thus initiating the electron bombardment mechanism, the temperature reaches 1700$^\circ$C within the IGISOL target chamber. The temperature increases linearly with the electron bombardment power as shown in Fig. \ref{fig: tp} but at present 1700$^\circ$C appears to be the usable limit. At higher temperatures the cavity becomes unstable and the worst scenario occurs when a runaway effect is seen in the bombardment current. The limit could be exceeded by using a current-stabilized power supply for the electron bombardment voltage. Applying more filament current does not yield any significant temperature gain as can be seen in Fig.\ref{fig: saturationc} and would only add an unnecessary ohmic load on the filament thus reducing its lifetime.


\subsection{Laser ion source performance}
\label{sec:7}
Initial tests of the hot cavity laser ion source were performed with metallic silver inserted into the cavity, using a filament heating current of $\sim$15 A and no electron bombardment in order to evaporate the silver and thus establish a proof of principle for the technique. The CVL was not used in these first tests and thus the final step ionization proceeded non-resonantly using photons from the two resonant steps and the fundamental output of a third Ti:Sapphire laser with a power of $\sim$1.6 W. In the following we describe the results obtained from laser wavelength scans, dipole magnet scans and extraction time measurements. 

\begin{figure}
	\centering
		 \includegraphics[width=8.5 cm]{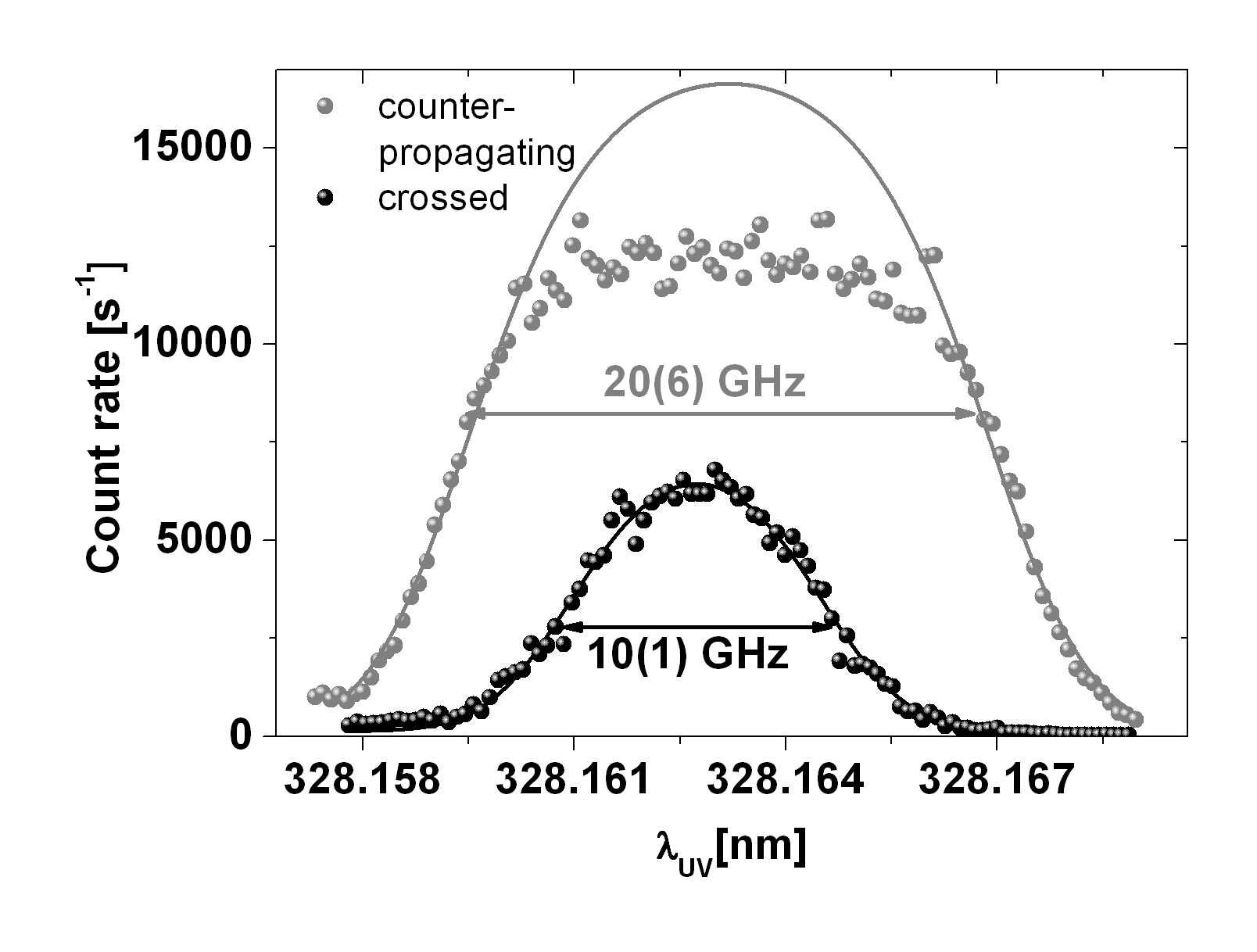}
	\caption{Laser wavelength scan for the UV excitation step in both crossed and counter-propagating geometries.}
	\label{fig: scan}
\end{figure}

Laser wavelength scans were performed over the first excitation step in order to study the differences between the two ionization modes, the crossed beams and counter-propagating mode.  A comparison between the two ionization geometries is illustrated in the wavelength scan of Fig. \ref{fig: scan}. The counter-propagating geometry results in a higher ionization efficiency, directly reflected by the increased count rate. Unfortunately, the ion count rate became saturated at the laser repetition rate (12 kHz was used rather than the usual 10 kHz). The reason for this is the coincidental arrival of ions originating from the same laser pulse on the detector being counted as a single event by the data acquisition system. Consequently, a direct comparison between the two ionization geometries is not possible. Recently, the system has been upgraded with an analog current readout to circumvent this issue.  To estimate the shape of the counter-propagating spectrum a fit was attempted where only the wings of the structure having a count rate of less than 10000 ions/s was taken into account. In this manner, a gain of a factor of two in the efficiency can be deduced compared to the crossed-beams geometry. However, it should be noted that due to the poor window transmission of the dipole separator magnet the realistic improvement in ionization efficiency may be significantly higher than estimated. The severe loss in laser power for all excitation steps due to the window transmission limits the use of the counter-propagating geometry to off-line experiments only. Due to the magnet design it is presently not possible to gain access to the window for cleaning. The reduction in the ionization efficiency in the crossed-beams geometry is compensated for by a factor of two improvement in the spectral resolution, ascribed to a reduction of the Doppler width of the transmission.

\begin{figure}
	\centering
		 \includegraphics[width=7.5 cm]{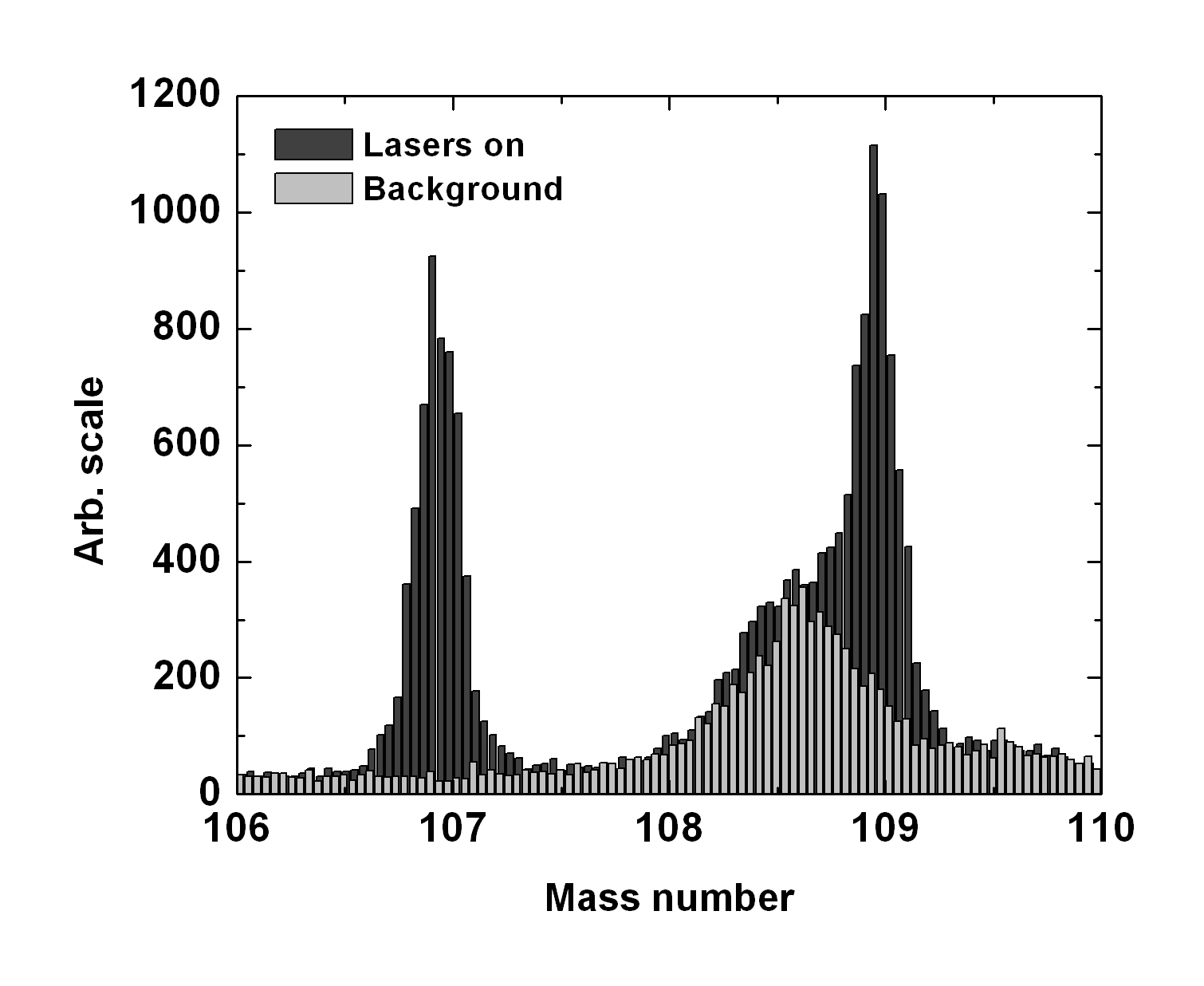}
	\caption{A dipole mass scan of stable $^{107}$Ag and $^{109}$Ag produced with the hot cavity laser ion source. The hot cavity - IGISOL separator has a mass resolving power of $\sim$470 deduced from the fit to the $^{107}$Ag peak. The oven produces a contaminant peak at \textit{A}=108.5 which distorts the natural abundances.}
	\label{fig: dipole}
\end{figure}

Stable silver consists of two isotopes $^{107}$Ag and $^{109}$Ag with natural abundances of 51.8\% and 	48.2\%, respectively. Fig. \ref{fig: dipole} shows a scan of the magnetic field of the IGISOL separator dipole magnet corresponding to the mass region of interest. When the lasers are off, a contaminant peak is seen, possibly surface-ionized material, at a mass number of 108.5 amu. This contaminant peak influences the apparent isotopic distribution seen in the mass scan with the lasers on. From a gaussian fit to the silver peak at \textit{A}=107 we have extracted a mass resolving power \textit{M}/$\Delta$\textit{M} $\sim$470. This is remarkably close to the typical mass resolving power achieved with the standard IGISOL front-end with the combination of the ion guide and rf sextupole extraction device \cite{karvonen08}.   

\begin{figure}
	\centering
		 \includegraphics[width=7.5 cm]{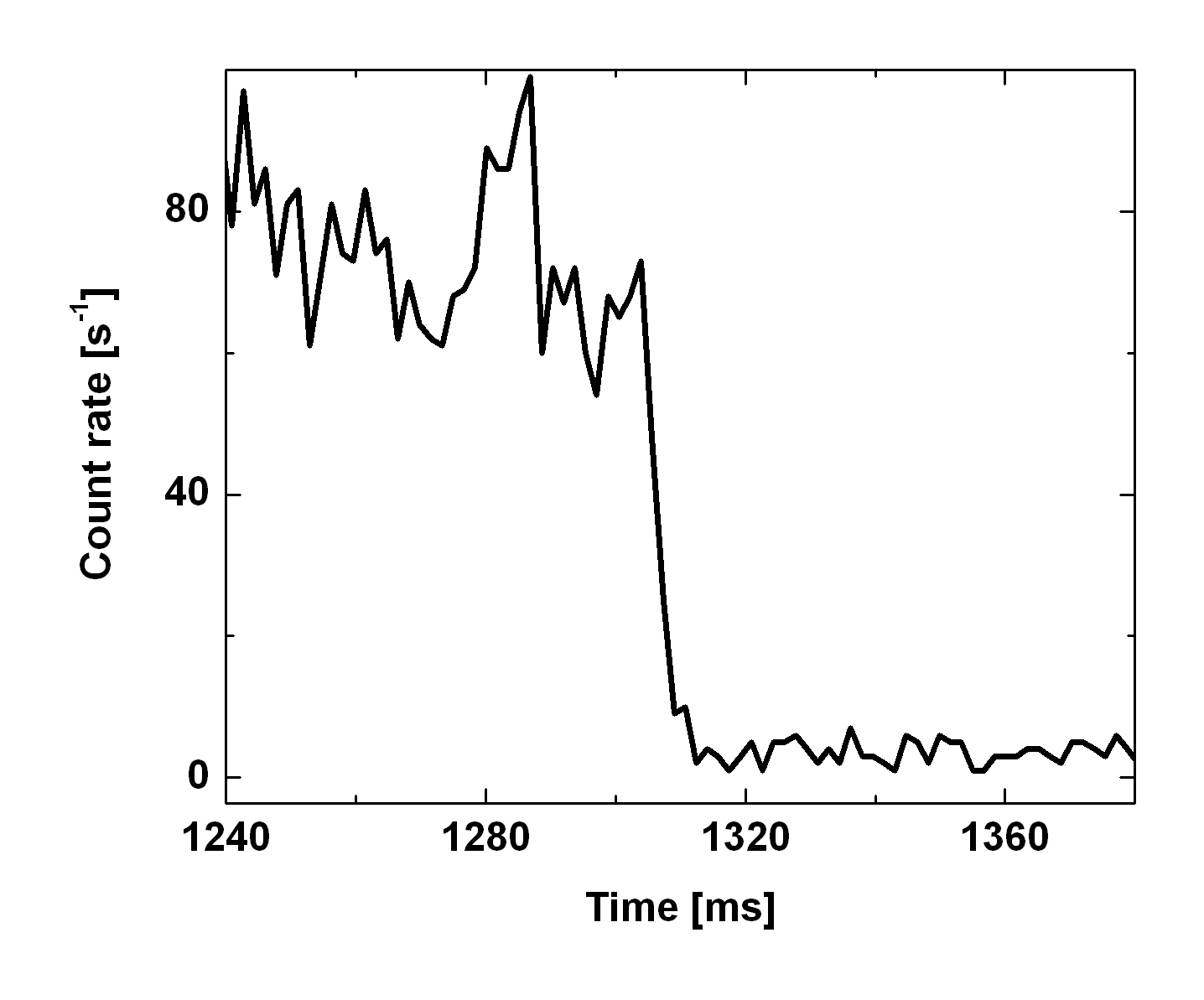}
	\caption{Time structure of $^{107}$Ag photo-ions.}
	\label{fig: extraction} 
\end{figure}

Finally, the origin of the ions was verified via a simple laser on-off test illustrated in Fig. \ref{fig: extraction}, followed by a fitting of the decay time of the resultant ion signal. A decay time of approximately 4 ms was determined which indicates that the ions originate from within the hot cavity rather than being surface-ionized \cite{kirch92}. A typical surface-ion time profile has a decay time of the order of $\mu$s \cite{kessler08}.

\section{Summary and outlook}

A hot cavity laser ion source has been successfully commissioned at the IGISOL separator facility, Jyväskylä. We have, for the first time, shown that the laser ionization process can be fully saturated, including the final non-resonant ionization step when a pump laser is used to directly ionize the atoms into the continuum. The FEBIAD-type hot cavity source used in this work has been fully characterized and is well-understood. The present system still has limitations however, the most important being the increasingly unstable electron bombardment current at catcher temperatures in excess of $\sim$1700$^\circ$C. In the work of Kircher \cite{kirch92} the catcher temperature of the FEBIAD-E source varied between 2200-2400$^\circ$C, and these values may in fact be exceeded with primary beam heating. The need to reach higher temperatures is likely to be extremely important for the diffusion of implanted silver recoil products from the graphite material. A new effort has been initiated between the authors of this paper and the JYFL ECR group in order to design and test a rf-based inductively-heated hot cavity which would replace the current FEBIAD design. It is expected that stable temperatures in excess of 2000$^\circ$C will be reached, based on current in-house experience with such devices \cite{RFO}.

First tests of the new inductively-heated cavity are expected to be performed early in 2009. A comparison with the present FEBIAD-type design will be made using the direct implantation and evaporation of a stable silver ion beam from the JYFL K-130 cyclotron. A test beam of 487 MeV $^{107}$Ag$^{21+}$ has been recently extracted from the cyclotron for precisely this application. A rotating nickel degrader foil of thickness 12.5 $\mu$m, similar to that used in recent stopped beam experiments in the laser ion guide \cite{moore08}, will be used to degrade the energy of the $^{107}$Ag beam such that the implantation depth closely matches that expected from the heavy-ion fusion-evaporation recoils of $^{94}$Ag. On success, the first on-line experiments to extract and selectively ionize radioactive silver isotopes/isomers will commence.

\section*{Acknowledgements}
We are indebted to the JYFL ECR group for the development of the Ag beam.
This work has been supported by the LASER Joint Research Activity project under the EU 6th. framework program ``Integrating Infrastructure Initiative-Transnational Access'', Contract number: 506065  (EURONS) and by the Academy of Finland under the Finnish Center of Excellence Program 2006-2011 (Nuclear and Accelerator Based Physics Program at JYFL).
%
%
%
%
%

\end{document}